\documentclass[twocolumn,prd,showpacs,floatfix]{revtex4}
\usepackage{graphicx}
\usepackage{dcolumn}
\usepackage{bm}

\def\A{{\bf A}}  

\def\x{{\bf x}}
\def\y{{\bf y}}
\def\k{{\bf k}}

\def\l{{\bf l}}
\def\q{{\bf q}}

\def\p{{\bf p}}

\def\SS{S^{(Q{\bar Q}G)}}

\def\lsim{\mathrel{\rlap{\lower4pt\hbox{\hskip1pt$\sim$}}
     \raise1pt\hbox{$<$}}}
\def\gsim{\mathrel{\rlap{\lower4pt\hbox{\hskip1pt$\sim$}}
     \raise1pt\hbox{$>$}}}

\newcommand{\gapproxeq}{\lower.7ex\hbox{$\;\stackrel{\textstyle>}{\sim}\;$}}
\newcommand{\lapproxeq}{\lower.7ex\hbox{$\;\stackrel{\textstyle<}{\sim}\;$}}

\begin{document}
\title{ Spontaneous chiral symmetry breaking in the linked cluster
  expansion } 

\author{Adam P. Szczepaniak and  Pawel Krupinski }

\address{
  Department of Physics and Nuclear Theory Center \\
  Indiana University, 
  Bloomington, Indiana   47405-4202 }

\begin{abstract}
We investigate dynamical chiral symmetry breaking 
 in the Coulomb gauge Hamiltonian QCD. Within the
  framework of the  linked cluster expansion we extend 
 the BCS ansatz for the vacuum and include 
  correlation beyond the quark-antiquark paring. In particular we 
 study the effects of the three-body correlations involving  
 quark-antiquark and transverse gluons. The high momentum behavior of
 the resulting gap equation is discussed and numerical  
 computation of the chiral symmetry breaking is presented.

\end{abstract}
\maketitle


\section{Introduction}
Chiral symmetry plays a major role in constraining the spectrum of 
 low energy QCD. At zero density it is spontaneously broken and the
  associated Goldstone bosons dominate the low energy, soft hadronic 
 interactions. The quark-gluon interactions which in vacuum break chiral
 symmetry may in dense matter, {\it e.g.} in the interior of neutron
 stars, lead to other, novel phases of the quark gluon plasma~\cite{cf1}. 
   The chiral properties of the QCD vacuum at 
 zero temperature and density have been  extensively studied in
 various approaches to soft  QCD~\cite{chsd,chha,Lis,Cotan,AA}. 
  In principle one could investigate it using lattice gauge
   methods. However, extrapolations of  lattice simulations 
  to small quark masses $m_{u,d} << 50-100 \mbox{ MeV}$ 
 (chiral extrapolation) still present a major challenge. 
 In approaches bases on a  Dyson-Schwinger formulation of QCD, 
  dynamical chiral symmetry breaking can be studied by analyzing the 
 behavior of the quark propagator.  Even though 
 no systematic truncation scheme of the Dyson series in QCD exists, 
   and in a majority of studies model interactions are 
  introduced, the approach gives a good description of the 
  low energy phenomenology.  
 In particular it enables to correctly predict many of the
  static properties of the low lying mesons and baryons, {\it i.e.} 
  masses and  charge moments, and simultaneously account  for the 
  dynamical chiral symmetry breaking as measured by the vacuum 
 expectation value of the scalar
  quark density, $\langle {\bar \psi} \psi \rangle \sim
  -(250\mbox{ MeV})^3$~\cite{Maris}. 
  This value follows from PCAC, Goldstone theorem and the current 
 algebra which result in the 
Gell-Mann-Oakes-Renner (or Thouless
  theorem) relation, 
 $-2 m_q \langle {\bar \psi} \psi \rangle = f_\pi^2 m_\pi^2. $
 Here, $m_q \sim 5-10\mbox{ MeV}$ is the current light quark mass,
 renormalized at the hadronic scale, 
  $f_\pi=93\mbox{ MeV}$ is the pion decay constant and $m_\pi$ is the 
 pion mass. Without explicit 
 chiral symmetry breaking $m_q=0$, the above relation cannot be used to 
 determine $\langle {\bar \psi}\psi \rangle$. However, as $m_q \to 0$ 
 no phase transition to a chirally symmetric state is expected,   
 and therefore the $~-(200-250\mbox{ MeV})^3$ should still be a 
 good estimate of the condensate in the chiral limit. 

 Spontaneous chiral symmetry breaking enables 
 to put the constituent quark representation of hadrons in a
  firm theoretical ground.  The bare 
  quark states defined with respect to the perturbative vacuum are 
  replaced by quasiparticle excitations  
  of the chirally noninvariant ground state. 
   Residual  interactions correlate the  quasiparticles 
 to form composite hadrons in which 
 each valence quasiparticle contributes kinetic energy of the order of a
  few hundred MeV. This is analogous to the constituent quark model
 representation of hadrons and therefore it might be possible 
 to further constraint the quark model phenomenology 
 from a first principle,  QCD 
 based analysis of dynamical chiral symmetry breaking. 
 Since the quark model picture calls for a Fock
 space representation it is most natural to consider a canonical,
 time-independent formulation of QCD. Coulomb gauge QCD offers such a
 framework~\cite{cl1,ss3,cl3}. 
 In the Coulomb gauge the single particle spectrum contains 
 only physical  degrees of freedom, {\it i.e.} 
  two transverse gluon polarizations. 
 As long as the gauge fields are restricted to the 
 fundamental modular region,  with no Gribov copies, the 
  Hamiltonian is positively defined, it leads to a
  continuous time evolution, and it is amenable to a variational treatment. 
  Finally the Coulomb gauge formulation leads to a natural realization
 of confinement. This arises because elimination 
  of the non-physical degrees of freedom
  through the gauge choice, ${\nabla \cdot \A} = 0$ results in an
 effective, long ranged instantaneous interaction between color
 charges. This interaction  is the analog of
  the Coulomb potential in QED. In QCD however, the colored Coulomb 
 gluons can  couple to transverse gluons leading to a Coulomb kernel
 which also  depends on the dynamical gluon degrees of freedom. As shown in 
 Ref.~\cite{ss7} 
 summation of the dominant IR contributions to the vacuum expectation
 value of the Coulomb operator results in a potential between color charges
  which grows linearly at large distances in agreement with
 lattice calculations~\cite{latt1}. In a self-consistent 
  treatment the same potential modifies the single gluon spectral
 properties and leads to an effective mass for 
  quasi-gluon excitations $~O(500-800\mbox{ GeV})$, 
  which is also in agreement with recent lattice
 calculations. The appearance of the gluon mass gap can 
 be  used to justify the implicit assumption of
  the quark model that mixing between valence quarks and Fock space
 sectors with explicit gluonic excitations is small. 
  We will return to this point in Section III. 

 The Coulomb gauge formulation provides a very natural
 starting point for building the constituent representation in accord
 with confinement and dynamical chiral symmetry breaking. 
 However, as it was noticed some time ago  in the Coulomb gauge 
 the simple  BCS treatment of the 
 vacuum is not sufficient to generate the right amount of chiral symmetry
 breaking. In particular if a pure linear potential is used, 
  $V(r) = b r $ with $b \sim 0.2-0.25\mbox{ GeV}^2$ as determined by 
  lattice calculations one typically obtains $|\langle {\bar \psi} 
 \psi \rangle|^{1/3} \sim 100\mbox{ MeV}$ {\it i.e.} too small by a
 factor of two~\cite{AA,chha,Cotan}. 
 The short range part of the Coulomb potential 
  requires proper handling of UV divergences and renormalization and
   in most recent studies has been ignored. As will be shown later, 
 it does significantly enhance the condensate and we  
  will  argue that the missing contribution can be accounted for 
  by three-particle correlations on top of the  BCS-like, particle-hole 
  vacuum. 

 The paper is organized as follows. In Section II we briefly discuss
 the canonical Coulomb gauge formalism and the 
  linked cluster expansion which enables to include multi-particle
 correlations into the many-body ground state. We will derive the 
 resulting contributions to the mass gap including up to three-body 
 correlations. The formalism is suitable for handling both 
 zero and finite density system and in this paper we will focus on the
 former. In Section III we discuss the approximations, numerical 
 results and possible sources of UV divergence and their 
 renormalization. Our conclusions and outlook are given in
Section IV.

\section{Coulomb gauge Hamiltonian and the linked cluster expansion}

 QCD canonically quantized in a physical gauge, {\it
  e.g.} Coulomb gauge, results in a Hamiltonian that can be represented in
  a complete Fock space defined by a set of single particle
   orbitals. One possibility is to choose the single particle basis
  as eigenstates of the kinetic (noninteracting) part of the
  full Hamiltonian, 
\begin{eqnarray}
H_0 & = &  H(g=0) =  \int d\x  \psi^{\dag}(\x)\left[
  -i\bbox{\alpha}\cdot \nabla + \beta m \right] \psi(\x)  \nonumber \\
 & &   + \int d\x \left[  \mbox{Tr } \bbox{\Pi}(\x)^2 + 
 \mbox{Tr } \left( {\bf \nabla} \times {\bf A}(\x) \right)^2 
  \right].  \label{h0} 
\end{eqnarray}
The vacuum, $|0\rangle$, 
  of $H_0$ is shown schematically in Fig.~1a.  The singe-particle 
 excitations at zero density correspond to adding gluons to the positive
  energy, parton-like levels and 
 quark-antiquark paris by creating a particle-hole excitation around
  the zero-energy Fermi surface. These excitations have energies given by,   
   $\epsilon^0_q(\k) = \epsilon^0_{\bar q}(\k) =
 \sqrt{m^2 + \k^2}$, $\epsilon^0_g(\k) = |\k|$ for  quarks, antiquarks 
 and gluons, respectively. 
 The quark fields in Eq.~(\ref{h0}) satisfy
 the canonical anticommutation relations and the gluons fields are given by 
  $\bbox{\Pi} \equiv \bbox{\Pi}^a
  T^a$ and ${\bf A} \equiv  {\bf A}^a T^a$
  and satisfy the canonical commutation relations for transverse fields, {\it
    i.e.} 
\begin{equation}
\left[ \bbox{\Pi}^a(\x), {\bf A}^b(\y) \right] = -i\delta^{ab} \bbox{\delta}_T(\nabla)
\delta^3(\x - \y),
\end{equation}
where $\bbox{\delta}_T(\nabla) = I - \bbox{\nabla}\otimes
\bbox{\nabla}/\bbox{\nabla}^2$. 
In terms of the single particle creation and annihilation operators,
 the color triplet of quark fields ($i=1,2,3$)  is given by, 
\begin{eqnarray}
\psi_i(\x) =  \sum_{\lambda=\pm1/2}\int {{d\k}\over {(2\pi)^3}} & & \left[ 
 u(\k,\lambda) b (\k,\lambda,i) \right. \nonumber \\
 & & \left.  + v(-\k,\lambda) d^{\dag}(
 -\k, \lambda,i) \right] e^{i\k\cdot\x}, \nonumber \\
\end{eqnarray}
where  $u$ and $v$ are solution of the free Dirac equation for a fermion
 with mass $m$. In the following we will restrict our discussion to chirally
  symmetric case {\it i.e.} from now on we will set $m=0$. 
The gluon field is given  by, 
\begin{eqnarray}
{\bf A}^a(\x) = \sum_{\lambda=\pm 1} \int {{d\k}\over {(2\pi)^3}}
 & &  {1\over \sqrt{2 \omega^0(\k)}}  \left[ 
 a(\k,\lambda,a) \bbox{\epsilon}(\k,\lambda) \right. \nonumber \\
& & \left.  + 
  a^{\dag}(-\k,\lambda,a) \bbox{\epsilon}^{*}(-\k,\lambda)  \right]
 e^{i\k\cdot\x},
 \nonumber \\
\end{eqnarray}
with $\omega^0(\k) = \epsilon^0_g(|\k|)$. 
The unrenormalized Coulomb operator is given by, 
\begin{equation}
H_C = {g^2 \over 2} \int d\x d\y \rho^a(\x) K_{ab}(\x,\y,{\bf A})
\rho^b(\y),
\end{equation}
where $\rho^a(\x) = \psi^{\dag}(\x) T^a \psi(\x) + 
 f^{abc} \bbox{\Pi}^b(\x)\cdot{\bf A}^c(\x)$ is the color charge density 
and the kernel $K$ is given by, 
\begin{equation}
K_{ab}(\x,\y,A) = \langle \x,a| {1\over {\bbox{\nabla}\cdot {\bf D}}}
( -\bbox{\nabla}^2)  {1\over {\bbox{\nabla}\cdot {\bf D}}} |\y,b\rangle,
\end{equation}
where  ${\bf D}$ is the covariant derivative in the adjoint
  representation, and the  $\langle {\bf x},a| \cdots |{\bf
  y},b\rangle$ matrix element  is given by 
 $\langle \x,a|{\bf D}|\y,b\rangle  = \left[ 
  \delta^{ab}\bbox{\nabla}_{\x} + g f^{acb} {\bf A}^c(\x)  
  \right] \delta^3(\x - \y)$, and  $\langle {\bf x},a|
  1/\bbox{\nabla}^2 |{\bf  y},b\rangle = -1/4\pi|{\bf x} - {\bf y}|$. 
  When $H_C$ is normal ordered with respect to the
  perturbative vacuum, $|0\rangle$ one might expect that the 
  mean field, Hartee-Fock corrections to the single particle
  energies could already generate an effective mass. 
  This is not the case.  
 The vacuum is a color singlet and thus the  direct 
   contribution from $H_C$ to a single fermion energy vanishes.  
  Furthermore, chiral symmetry of the Hamiltonian and of the 
 perturbative, $| 0\rangle$ vacuum protects the exchange term from mass 
 generation. The effective mass can only be  
  obtained if quark-antiquark correlations are introduced into the
  ground state as shown schematically in Fig.~1b. 
\begin{figure}[htb]
\includegraphics[scale=0.4]{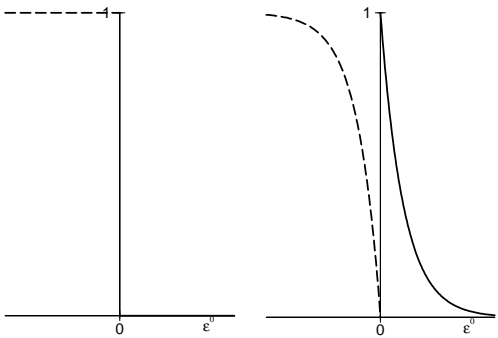}
\caption{Schematic representation of the particle, 
 $n^+ \equiv \langle b^{\dag} b \rangle$ (solid) and hole 
  $n^- \equiv 1 - \langle d^{\dag} d \rangle $ (dashed)  occupations   in 
  partonic (left)  and BCS (right) ground state }
\end{figure}

The full, unrenormalized Coulomb gauge Hamiltonian has the following 
structure~\cite{cl3,ss7,CL,swif}, 
\begin{equation}
H = H_0 + H_C + V_{qg} + V_{3g} + V_{4g} + H_{corr}.
\end{equation}
Here $V_{qg}$ is the quark-transverse gluon interaction, 
\begin{equation}
V_{qg} = g \int d\x \psi^{\dag}(\x) \bbox{\alpha}\cdot {\bf A} \psi(\x),
\end{equation}
 and  $V_{3g}$ and $V_{4g}$ represent 3-- and 4-- transverse gluon couplings
 arising from the nonabelian part of the magnetic field, 
  ${\bf B}^a = \nabla \times {\bf A}^a  +  g f^{abc} {\bf A}^b(\x)
  \times {\bf A}^c(\x)$. Finally $H_{corr}$ contains terms which come 
  from a commutator of the determinant of the Faddeev-Popov operator, 
 ${\cal J} = Det({\bbox{\nabla}\cdot {\bf D}})$ and the gluon
 canonical momentum $\bbox{\Pi}$. The detailed analysis of this
 Hamiltonian, emergence of confinement and 
  issues related to renormalization in the gluon sector were
  discussed in Ref.~\cite{ss7}. 

\subsection{ Linked cluster expansion} 

 Since the Fock space basis generated by the set of single  
 particle creation operators, $b^{\dag}$, $d^{\dag}$, $a^{\dag}$ 
 is complete, the true ground
 state,  $|\Omega\rangle$ of $H$ can be written as,  
\begin{widetext}
\begin{equation}
|\Omega\rangle =  \left[  1 + \sum_{1 2} 
 F^{(q{\bar q})}_{ 1 2}b^{\dag}_1 d^{\dag}_2 + 
\sum_{12} F^{(gg)}_{ 1 2}a^{\dag}_1 a^{\dag}_2 + 
 \sum_{1 2 3} F^{(q{\bar q}g)}_{1 2 3} b^{\dag}_1 d^{\dag}_2 a^{\dag}_3 
   +   
\sum_{1 2 3 4} F^{(q{\bar q}q{\bar q})}_{1 2 3 4 } b^{\dag}_1 d^{\dag}_2
  b^{\dag}_3 d^{\dag}_4 + \cdots \right] |0\rangle. \label{Fe}
\end{equation}
\end{widetext}
Here $F^{(n)}_{1 2 \cdots n}$ represent wave functions of 
 n-body clusters in the vacuum, and $1,2\cdots$ collectively denote 
 quantum numbers of single particle orbitals. 
 This expansion is however impractical since it does not
differentiate between connected (linked) and disconnected
 contributions. For example, at the
 2-quark-2-antiquark level there are disconnected contributions 
 of the  type, $F^{(q{\bar q}q{\bar q})}_{1234} = F^{(q{\bar q})}_{12}
 F^{(q{\bar q})}_{34}$, {\it i.e.} part of the 
 $n$-particle cluster contribution 
 originates from products of smaller, $m<n$, $m$-particle 
 clusters. 

 The essence of the linked cluster expansion is based on the observation that 
 all multi-particle correlation in the ground state, including the
 disconnected ones can be accounted for by proper resummation of the 
 linked clusters only. This is achieved by 
 writing the full ground state as~\cite{exps}  
\begin{equation}
|\Omega\rangle = e^{-S}|0\rangle,
\end{equation}
with $S$ having the expansion
\begin{widetext}
\begin{equation}
S = \sum_n S^{(n)} = 
 \sum_{1 2} S^{(q{\bar q})}_{ 1 2}b^{\dag}_1 d^{\dag}_2
+  \sum_{1 2} S^{(gg)}_{ 1 2}a^{\dag}_1 a^{\dag}_2
 +  \sum_{1 2 3} S^{(q{\bar q}g)}_{1 2 3} b^{\dag}_1 d^{\dag}_2 a^{\dag}_3
 +   \sum_{1 2 3 4} S^{(q{\bar q}q{\bar q})}_{1 2 3 4 } b^{\dag}_1 d^{\dag}_2
  b^{\dag}_3 d^{\dag}_4 + \cdots,  \label{s}
\end{equation}
\end{widetext}
 with the operators $S$ including connected pieces only. 
 Comparing Eq.~(\ref{Fe}) and Eq.~(\ref{s}) we find for example that, 
\begin{eqnarray}
& & F^{(q{\bar q})}_{12} = S^{(q{\bar q})}_{12},  \;\; 
F^{(q{\bar q}g)}_{123} = S^{(q{\bar q}g)}_{123}, \;\;  \nonumber \\
& & F^{(q{\bar q}q{\bar q})}_{1234} = S^{(q{\bar q}q{\bar q})}_{1234} 
+ {1\over 2} S^{(q{\bar q})}_{12} S^{(q{\bar q})}_{34}, \cdots 
\end{eqnarray}
{\it i.e.} the general expansion of Eq.~(\ref{Fe}) is obtained  with all 
 disconnected contributions constrained by the connected ones. 
  The expansion coefficients, $S^{(n)}_{12\cdots n}$ 
 can be determined from the eigenvalue equation for $|\Omega\rangle$, 
\begin{equation}
 e^{S} H e^{-S} |0\rangle = E_\Omega |0\rangle. 
\end{equation}
This equation projected onto the partonic Fock space basis leads to a 
 set of equations 
 for the amplitudes $S^{(n)}_{12\cdots n}$ and the ground state energy,
 $E_\Omega$,
\begin{widetext}
\begin{equation}
\langle 0| e^{S} H e^{-S} |0\rangle = E_\Omega, \;\; 
\langle q_1,q_2,\cdots q_{n_q};{\bar q}_{1'},{\bar q}_{2'},\cdots {\bar
 q}_{n_{\bar q}}; 
 g_{1''},g_{2''},\cdots g_{n_g}| e^{S} H e^{-S}   |0\rangle = 0,
 n_q, n_{\bar q},  n_g = 1,2,\cdots . \label{s1}
\end{equation}
\end{widetext}
In  a  nonrelativistic many-body system 
 the  Hamiltonian is typically a  polynomial in the field operators. Since 
 each  $S^{(n)}$ contains only particle creation operators, the 
 matrix elements of $e^{S} H e^{-S}$ between
 an $n$-particle state and the free vacuum will involve only a 
 finite number of terms arising from the expansion of the
 exponentials. For example in a 
 typical case when $H = H_0 + V$ with $H_0$ being a one body ({\it
 e.g.} kinetic) operator
 and $V$ a two-body potential one has, 
\begin{equation}
e^{S} H e^{-S} = H + [S,H] + \cdots + {1\over {4!}}[S,[S,[S,[S,H]]]].
\end{equation}
In this case an approximation to Eq.~(\ref{s1}), is 
 fully  specified by a number of clusters retained in $S$. 
 This is, however, not the case for the relativistic system discussed here. 
 The expansion of the Coulomb kernel leads to an infinite series of
  operators to all orders  in the transverse gluon field. 
 Thus an approximation to Eq.~(\ref{s1})  consists of 
  specifying which clusters are kept in the definition of $S$ 
   and of a truncation scheme in evaluation of matrix
   elements of $e^{S} H e^{-S}$.

  The truncation of $S$ limits the number of quark-antiquark-gluon 
 correlations build into the ansatz for the ground state. At first one
   might think that such a truncation would be hard to justify since 
 any hadronic state, including the vacuum should have a large
 (infinite) number of partons. However, the first two terms in $S$, $S^{(q{\bar q})}$ and $S^{(gg)}$  
 change the single particle excitation spectrum and effectively 
 replace the partonic basis by that of massive quasiparticles. This
 is known as the Thouless reparameterization~\cite{BR} 
 and is equivalent to the 
  BCS ansatz for the vacuum which contains 
 two-body, quark-antiquark and gluon-gluon correlations. 
 The BCS ansatz leads to
   the chiral gap, constituent mass for the  quarks as well as 
 effective mass for the  transverse gluons.  Iterative contributions of 
  multiparticle states which determine the wave functions of larger
   clusters, $S^{(n)}$, $n > 2$ are therefore 
  suppressed by the quasiparticle energy gap. This gap is 
   $O(400-600 \mbox{ MeV})$ for quark-antiquark excitations and  $O(500\mbox{
    MeV}- 800 \mbox{ GeV})$ for a gluonic excitation. The former follows
 from the typical constituent quark mass and the later 
   from the gluon spectrum in a presence of static color sources as 
 calculated on the lattice~\cite{latt1} and are consistent with explicit
 calculation using the BCS gluonic ansatz for the
 Hamiltonian~\cite{ss7}. 
 The transformation form the partonic to the quasiparticle
    basis, generated by  $S^{(2)}$, proceeds as follows.  
  The (unnormalized) 
  quasiparticle, BCS  vacuum $|\Omega_{BCS} \rangle$ is defined as, 
\begin{equation} 
|\Omega_{BCS} \rangle \equiv e^{-S^{(2)}} |0 \rangle, 
\end{equation}
with, 
\begin{eqnarray}
S^{(2)}  = S^{(q{\bar q})} + S^{(gg)} = & & 
  \sum_{12} S^{(q{\bar q})}_{12} b^{\dag}_1 d^{\dag}_2  \nonumber \\
& &  + \sum_{12} S^{(gg)}_{12} a^{\dag}_1 a^{\dag}_2, 
\end{eqnarray}
so that 
\begin{equation}
|\Omega \rangle = e^{-\sum_{n>2} S^{(n)} } |\Omega_{BCS} \rangle.
\end{equation}
A canonical transformation which maps the set of free particle
 operators $b,b^{\dag},d,d^{\dag},a,a^{\dag}$ onto a set of
 quasiparticle operators $B,B^{\dag},D,D^{\dag},\alpha,\alpha^{\dag}$ 
 is defined by 
\begin{eqnarray}
& & B_1 = {1\over {\sqrt{ 1 + |S^{(q{\bar q})}|^2 }}} b_1  
+ \sum_{2} {{ S^{(q{\bar q})}_{12} }\over {\sqrt{ 1 + |S^{(q{\bar q})}|^2 }}}
 d^{\dag}_2, \nonumber \\
& & D_1 = {1\over {\sqrt{ 1 + |S^{(q{\bar q})}|^2 }}} d_1  
- \sum_{2} b^{\dag}_2 {{ S^{(q{\bar q})}_{21} }\over {\sqrt{ 1 +
      |S^{(q{\bar q})}|^2 }}},
\nonumber \\
& & \alpha_1 = {1\over {\sqrt{ 1 - |S^{(gg)}|^2 }}} a_1
+ \sum_{2} {{ S^{(gg)}_{12} }\over {\sqrt{ 1 - |S^{(gg)}|^2 }}}
 a^{\dag}_2, \label{BCS}
\end{eqnarray}
where $ |S^{(q{\bar q})}|^2\delta_{12} \equiv  \left [ S^{(q{\bar q})} 
 {S^{(q{\bar q})}}^{\dag}
\right]_{12}$ and similarly for $|S^{(gg)}|$.  These  
 quasiparticle operators satisfy the canonical
(anti)commutation relations, they annihilate the BCS ground state,   
\begin{equation}
 B_1 |\Omega_{BCS} \rangle = D_1 |\Omega_{BCS}\rangle = \alpha_1
|\Omega_{BCS} \rangle = 0,
\end{equation}
and generate a complete Fock space. The 
  eigenvalue conditions for the 
 vacuum, Eq~(\ref{s1}) can therefore be rewritten in the
  quasiparticle basis, 
\begin{widetext}
\begin{eqnarray}
{ { \langle \Omega_{BCS} | e^{S} H e^{-S} |\Omega_{BCS} \rangle} 
 \over { \langle \Omega_{BCS} | \Omega_{BCS} \rangle }} = E_\Omega, \;\; 
\langle Q_1, Q_2, \cdots Q_{n_Q};{\bar Q}_{1'},{\bar Q}_{2'},\cdots
 {\bar Q}_{n_{\bar Q}}; 
 G_{1''},G_{2''},& & \cdots G_{n_G}|  e^{S} H e^{-S} |\Omega_{BCS} 
 \rangle = 0,  \nonumber \\ 
& & n_Q, n_{\bar Q}, n_G = 1,2,\cdots .\label{S1}
\end{eqnarray}
\end{widetext}
Here the operator $S$ contains contributions from 3-quasiparticle
cluster and higher,
\begin{equation}
S = \sum_{123} {\tilde S}^{(Q{\bar Q} G )}_{123} B^{\dag}_1 D^{\dag}_2
\alpha^{\dag}_3 + \cdots . 
\end{equation}
The matrix elements ${\tilde S}^{(n)}_{12\cdots n}$ can be related 
to $S^{(n)}_{12\cdots n}$ by replacing the free particle operators 
  by the quasiparticle 
operators. From the structure of Eq.~(\ref{BCS}) it follows that for
given $n$ the operators ${\tilde S}^{(n)}$ are a linear 
   combination of $S^{(i)}$ including  $i \le n$. 
   Since Eq.~(\ref{BCS}) defines a canonical
 transformation the two sets of equations, Eq~(\ref{s1}) and
Eq.~(\ref{S1}) are equivalent and one can simply use the 
 later {\it i.e.} work directly in the quasiparticle basis without
 referring to the partonic basis. As suggested by the quark model 
  it is preferred to represent  
  low energy QCD eigenstates in terms of quasiparticle,  
 quark and gluon excitations. 
 From now on we will work the matrix elements of ${\tilde S}$ 
 in the quasiparticle basis and for simplicity rename them as ${\tilde
   S}^{(n)} \to S^{(n)} $. 

  As mentioned earlier, in QCD, with $S = \sum_n S^{(n)}$ truncated at some
  maximal $n$, Eq.~(\ref{S1}) still contain an infinite 
 number of terms arising from the expansion of $e^{S} H e^{-S}$.  
 Since this (infinite) series is related to the  multi-gluon 
  structure of the Coulomb operator, $K(\x,\y,{\bf A})$, 
  it can be organized according 
  to how each of the terms  renormalizes the 0-th order Coulomb potential, 
 $K_{ab}(\x,\y,0) = \delta_{ab}/4\pi|\x-\y|$. To illustrate this
 consider truncating $S$ at $n=2$. The {\it lhs.} of the first
  equation in ~(\ref{S1}) reduces to the expectation value of $H$ in
  the BCS vacuum,
\begin{equation}
\langle \Omega_{BCS} |e^{S} H e^{-S} |\Omega_{BCS} \rangle  = \langle \Omega_{BCS}
|H| \Omega_{BCS} \rangle.
\label{ex}
 \end{equation}
 The lowest order (in the loop
 expansion) diagrams are shown on the left side of  Fig.~2. 
   The matrix element  $\langle \Omega_{BCS} 
|H| \Omega_{BCS} \rangle $ 
 defines an effective potential $V_{eff}(\x-\y)$, by
\begin{equation}
V_{eff}(\x - \y) \equiv g^2 {{\delta_{ab}}\over {N_c^2 - 1}} 
 \langle \Omega_{BCS} | K_{ab}(\x,\y,{\bf A})  | \Omega_{BCS}
 \rangle. \label{veffbcs}
\end{equation}
It is straightforward to identify diagrams which give the dominant
 contribution to $V_{eff}$ in both, the IR ($|\x-\y| >>
 1/\Lambda_{QCD}$) and the UV
 ($|\x-\y|<<1/\Lambda_{QCD}$). 
 In the IR region these are given by diagrams which, at
  a given loop order contain the maximum number of soft potential,
 $K(\x,\y,0)$ lines; the UV region is 
 dominated by loops with the smallest number of vertices. 
   The series of ring and rainbow diagrams, shown in
 Fig.~2, accounts for the leading IR and UV contributions to
 $V_{eff}$, respectively. The 
 approximation can be systematically improved by taking into account
 the subleading contributions {\it e.g.} vertex renormalization~\cite{ss7,swif}
\begin{figure}[htb]
\includegraphics[scale=0.4]{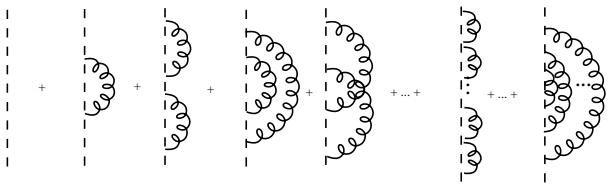}
\caption{\label{fig:2} 
A few lowest order contributions to $V_{eff}$. The two
  right most diagrams represent diagrams in the ring and rainbow 
  series.  The dashed line stands for the 0-th order Coulomb potential
  $g^2 K_0 = g^2/4\pi|\x-\y|$. }
\end{figure}
 If larger clusters in $S$ are retained, 
 the expansion of  $e^{S} H e^{-S}$ generates operators that
 have nonvanishing matrix elements between the vacuum and states
 with an arbitrary large number of particles. This is because as long
  as $S^{(n)}$ contains a gluon operator an infinite number of commutators,
 $[S^{(n)},[S^{(n)},[ \cdots [S^{(n)},H] \cdots]]]$ are nonvanishing.   
  Their contribution arise from contracting gluons 
 from each $S^{(n)}$ with gluons from the Coulomb operator. For
 example a 
   term in $S$ which contains pure glue operators (no quark or 
 antiquark)  will contribute 
  to any matrix element in Eq.~(\ref{S1}) with any 
 number of particles (gluons). 
 This is illustrated in Fig.~3 for $S^{(3g)}$. It is clear, however, 
  that this type of corrections have the effect  of simply  
 renormalizing $V_{eff}$, beyond the BCS-like contributions shown in Fig.~2. 
 Since the operators $S^{(n)}$ commute with each other one possibility
 is to consider the effects of the pure gluon operators first,
 generate the new  effective interaction and then introduce
 clusters which contain quark and antiquark operators. 
 Since $H$ is a finite order polynomial in the quark
 operators, each term in $S$ containing only quark and antiquark 
 operators will lead to a finite number 
 of terms in a matrix element between the BCS vacuum and a 
 multiparticle state with a fixed, $n=n_Q + n_{\bar Q} + n_G$.  
\begin{figure}[htb]
\includegraphics[scale=0.3]{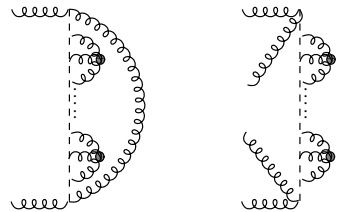}
\caption{\label{fig:3} Example of contributions to 
 $[S^{(n)},[S^{(n)},[ \cdots [S^{(n)},H] \cdots]]]$, for $S=S^{ggg}$ 
 and $H = H_C$. The two diagrams contribute to a matrix element with
 $n_G=2$ and $n_G=4$ respectively. Here the dashed line represents
 the Coulomb potential dressed by the BCS corrections,
 Eq.~(\ref{veffbcs}), {\it e.g.} through resummation of the ring-rainbow
 series shown in Fig.~2. }
\end{figure}

To summarize, the linked cluster expansion of the QCD ground state is
 much more  complicated that in a typical nonrelativistic many-body 
 problem. Nevertheless it can be used to systematically improve the
 BCS approximation. It is important to notice, however, that even the 
 BCS ground state already  probes the nonabelian multi-gluon dynamics 
 via $\langle \Omega_{BCS}|
 K(\x,\y,\A) | \Omega_{BCS} \rangle$. In BCS this leads to and
 effective  interaction  which is very close to the potential between
 color sources and when treated selfconsistently 
 leads to a quasiparticle (constituent) representations. 

\subsection{ $Q {\bar Q} G$ contribution to the quark mass gap } 

 In the following we will concentrate on the dynamical chiral symmetry
 breaking and therefore consider vacuum properties in 
 quark sector. 
\begin{figure}[htb]
\includegraphics[scale=0.44]{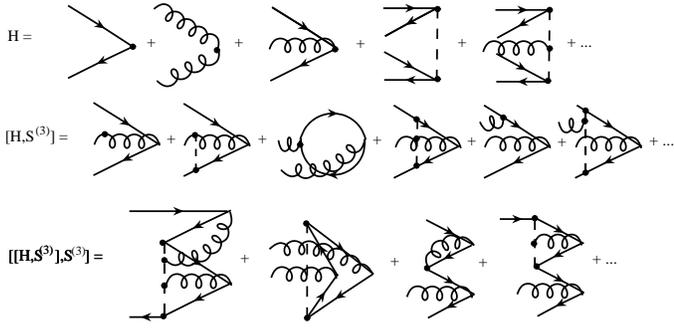} 
\caption{\label{fig:4} Operators from Eq.~\ref{2} which contribute 
 to matrix  elements $\langle n_Q,n_{\bar Q},n_G| \cdots |\Omega_{BCS} \rangle$ 
 for $n_Q \le 1$, $n_{\bar Q} \le 1$ and $n_G \le 2$  
  As in Fig.~3, the 
 potential (dashed) line is the $V_{eff}$ of Eq.~(\ref{veffbcs}). The
 matrix element, $\SS$ corresponds to the rightmost vertices. }
\end{figure}
 As mentioned earlier the BCS mechanism of quark-antiquark
 pairing seems to be insufficient to account for the full dynamical
 symmetry breaking. 
 We will discuss this point quantitatively in the
 following section. 
Our interest here is in extending the 
 BCS approximation by including the effects of the next to
  leading (beyond BCS) order in the cluster expansion {\it i.e.} 
  the 3--particle cluster contribution to the vacuum. 
 We will therefore study, 
\begin{equation}
S \to \SS = \sum_{123} \SS_{123} B^{\dag}_1
  D^{\dag}_2 \alpha^{\dag}_3. \label{gg1}
\end{equation}
The quark gap equation follows from, 
\begin{equation}
\langle Q_1 {\bar Q}_2 | e^{\SS} H e^{-\SS} | \Omega_{BCS} \rangle =
0. \label{1}
\end{equation}
This equation determines single particle orbitals and
therefore it also  gives the 
 quasiparticle spectrum via, $\epsilon_1\delta_{12} =
\langle Q_1|H|Q_2\rangle $. There is a finite
 number of terms contributing  to, Eq.~(\ref{1})
\begin{widetext}
\begin{equation}
0 = \langle Q_1 {\bar Q}_2 | H  + [\SS,H] + {1\over
  {2!}}[\SS,[\SS,H]]  +
{1\over {3!}} [\SS,[\SS,[\SS,H]] |
\Omega_{BCS}\rangle .\label{2}
\end{equation}
\end{widetext}
The series is finite because starting at 
 $O(\left[{\SS}\right]^4 \sim \left[B^{\dag} D^{\dag}\right]^4 )$
 commutators will produce operators which have at least
 2-quark and 2-antiquark creation operators and these  vanish between
  $\langle Q{\bar Q}|$ and $|\Omega_{BCS} \rangle$. Some of the
 contributions to Eqs.~(\ref{2}) and  ~(\ref{3}) are shown in Fig.~4.
In order to solve Eq.~(\ref{2}) and determine the single particle
 basis, it is necessary to first solve for the amplitude $\SS$.
  This amplitude can be obtained by projecting $e^{\SS} H e^{-\SS}$
 onto the three particle cluster, 
\begin{widetext}
\begin{equation}
\langle Q_1 {\bar Q}_2 G_3 | H   + [\SS,H] + {1\over {2!}}[\SS,[\SS,H]]  
 + {1\over {3!}}[\SS,[\SS,[\SS,H]]]
|\Omega_{BCS} \rangle = 0, \label{3}
\end{equation}
\end{widetext}
which also contains a finite number of terms. The two 
equations Eq.~(\ref{2}) and Eq.~(\ref{3}) form a set of coupled nonlinear,
 integral equations for the amplitude $\SS$ and the single particle
 orbitals (or the BCS angle, Eq.~(\ref{BCS})). In this paper we will 
   simplify these equations by linearizing them with respect to
  $\SS$,  Eq.(~\ref{3}) then yields, 
\begin{equation}
 \SS | \Omega_{BCS} \rangle = 
\sum_n | n \rangle {1\over {E_n - E_{\Omega_{BCS}} }}
 \langle n | H | \Omega_{BCS} \rangle 
\end{equation}
Here $|n\rangle$ is the set of eigenstates of $H$ in the three
 particle $Q{\bar Q} G$ subspace, 
 \begin{equation}
H |n \rangle = (E_n - E_{\Omega_{BCS}}) |n\rangle.
\end{equation}
The contribution from  $\SS$ to the quark gap in Eq.~(\ref{2}) 
 is then given by, 
\begin{eqnarray}
\delta_{12,\Omega} \delta m_g & \equiv &   \langle Q_1 {\bar Q}_2 | 
 [\SS,H] |\Omega_{BCS} \rangle \nonumber \\
& = & 
- \sum_n \langle Q_1 {\bar Q}_2| H | n  \rangle
 {1\over {E_n - E_{\Omega_{BCS}}}}
 \langle n | H | \Omega_{BCS} \rangle.   \nonumber \\
\end{eqnarray}
Here, $\delta_{12,\Omega}$ symbolizes the product of all $\delta$-functions 
 which restrict the quantum numbers of 
   $|Q_1{\bar Q}_2 \rangle$ to be same as of the vacuum. 
  With inclusion of $\delta m_g$ the gap equation can be
written as,
\begin{equation}
0 = \delta m_0 + \delta m_C + \delta m_g, \label{gap}
\end{equation}
where the BCS part given by 
\begin{eqnarray}
\delta_{12,\Omega} \left[ \delta m_0 + \delta m_C \right] 
&  = &  \langle Q_1{\bar Q}_2| H | \Omega_{BCS} \rangle \nonumber \\
& = & 
\langle Q_1 {\bar Q}_2 | H_0 + H_C | \Omega_{BCS} \rangle .
\end{eqnarray}
The three contributions to the gap equation are illustrated in Fig.~5.  
\begin{figure}[htb]
\includegraphics[scale=0.45]{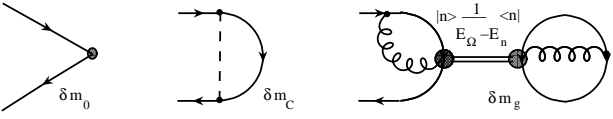} 
\caption{\label{fig:5} The three contributions to Eq.~(\ref{gap}). $\delta
  m_0$ is determined by the kinetic term, $\delta m_C$ by $V_{eff}$ 
  and $\delta m_g$ is the contribution of the $Q{\bar Q} G$  cluster.}
\end{figure}
In the next section we will write down the explicit form of the gap
equation and discuss the numerical solution. 

\section{Quark mass gap}

From translational, rotational and global color 
 invariance of the vacuum it follows that for each quark flavor, 
\begin{widetext}
\begin{equation}
\sum_{12} S^{(q{\bar q})}_{12} b^{\dag}_1 d^{\dag}_2  =  
\sum_{\lambda_q\lambda_{\bar q},i_qi_{\bar q}}\int {{d\k}\over {(2\pi)^3}} S^{(q{\bar q})}(|\k|) b^{\dag}(\k,\lambda_q,i_q)
 \left[ \bbox{\sigma}\cdot{\hat\k} \right]_{\lambda_q,\lambda_{\bar
 q}} \delta_{i_q,i_{\bar q}}
  d^{\dag}(-\k,\lambda_{\bar q},i_{\bar q}),
\end{equation}
\end{widetext}
 The chiral angle is given by ({\it cf.} Eqs.~(\ref{BCS})),
\begin{equation}
\tan_\k \equiv \tan\phi_q(|\k|) = {{2S^{(q{\bar q})}}\over 
 {1 - |S^{q{\bar q}}|^2}} = {{2S^{qq}(|\k|)}\over {1 - (S^{qq}(|\k|))^2}}.
\end{equation}
 To evaluate the matrix elements in Eq.~(\ref{S1}) the 
 Hamiltonian 
 needs to be expressed in terms of the quasiparticle operators. This 
 can simply be done by noticing that in the quasiparticle basis 
 the field operators become,  
\begin{eqnarray}
\psi_i(\x) = \sum_{\lambda=\pm1/2}\int {{d\k}\over {(2\pi)^3}} & & \left[ 
 U(\k,\lambda) B (\k,\lambda,i) \right. \nonumber \\
 & & \left. + V(-\k,\lambda) D^{\dag}(
 -\k, \lambda,i) \right] e^{i\k\cdot\x}, \nonumber \\
\end{eqnarray}
 where the quasiparticle spinors $U$ and $V$ are given by 
\begin{eqnarray}
& & U(\k,\lambda) = {1\over \sqrt{2E(E+M)}} 
\left( \begin{array}{c} (E+M)\chi(\lambda) \\
                 A\bbox{\sigma}\cdot \hat{\k} \chi(\lambda)
 \end{array} \right), \nonumber \\
& & V(-\k,\lambda) = {1\over \sqrt{2E(E+M)}} 
\left( \begin{array}{c}  -A\bbox{\sigma}\cdot \hat{\k} \chi(\lambda)\\
      (E+M)\chi(\lambda) \end{array} \right)
\end{eqnarray}
with $E=E(|\k|)$, $M=M(|\k|) = E \sin_\k$, $A=A(|\k|) = E\cos_\k$. 
 Here we have 
 introduced an arbitrary function $E(\k)$ to make the expression for
  the single quasiparticle 
  wave functions analogous to those of free particles, but it is
  clear that $U$ and $V$ do not depend on $E$ but only on the chiral
 angle.   

Similarly for the gluon fields we have, 
\begin{eqnarray}
& &\sum_{12}S^{(gg)}_{12}  a^{\dag}_1 a^{\dag}_2  =   \nonumber \\
& & \sum_{\lambda,a} 
 \int {{d\k}\over {(2\pi)^3}} S^{(gg)}(|\k|) a^{\dag}(\k,\lambda,a)
  a^{\dag}(-\k,\lambda,a). 
\end{eqnarray} 
and in terms of the quasi-gluon operators  the fields are given by, 
\begin{eqnarray}
{\bf A}^a(\x) & = &  \sum_{\lambda=\pm 1} \int {{d\k}\over {(2\pi)^3}}
  {1\over \sqrt{2 \omega(|\k|)}} \left[ 
 a(\k,\lambda,a) \bbox{\epsilon}(\k,\lambda) \right. \nonumber \\
& & \left. + 
  a^{\dag}(-\k,\lambda,a) \bbox{\epsilon}^{*}(-\k,\lambda)  \right]
 e^{i\k\cdot\x}, 
\end{eqnarray}
with
\begin{equation}
\omega(|\k|) = |\k| (\cosh_\k + \sinh_\k), 
\end{equation}
and 
\begin{equation}
\tanh_\k = \tanh \phi_g(|\k|) =  {{2S^{(gg)}}\over 
 {1 + |S^{gg}|^2}} = {{2S^{gg}(|\k|)}\over {1 + (S^{gg}(|\k|)^2}}.
\end{equation}
Truncating $S$ at the ${\bar Q}QG$ level leads to uncoupled gluon and
quark gap equations. The gluon gap equation was studied in
Ref.~\cite{ss7}. The gluon gap function $\omega(|\k|)$ was determined
by the matrix element of the Coulomb operator in the BCS vacuum, which
in turn was selfconsistently determined by the gluon mass gap. It was  found
 that a good analytical approximation to, $V_{eff}(\x-\y)$ ({\it c.f.}
Eq.~(\ref{veffbcs}) ) is, in momentum space, given by,  
\begin{equation}
V_{eff}(\k) = {{f(\k) d^2(\k)}\over \k^2},   \label{potss}
\end{equation} 
where $d(\k)$ is the expectation value
of the Faddeev-Popov operator and it is approximately given by, 
\begin{equation}
d(\k) = \left\{ \begin{array}{cc} 3.5 \left( {m_g\over |\k|} \right)^{0.48}
       & \mbox{for } |\k| < m_g \\
  3.5 \left( {{ \log(2.41)}\over {\log(1.41 + |\k|^2/m_g^2)}}
\right)^{0.4} & \mbox{for } |\k| > m_g 
\end{array}  \right. , \label{d}
\end{equation}
and
\begin{equation} 
f(|\k|) = \left\{ \begin{array}{cc}  \left( {m_g\over |\k|} \right)^{0.97}
       & \mbox{for } |\k| < m_g \\
   \left( {{ \log(1.82)}\over {\log(0.82 + |\k|^2/m_g^2)}}
\right)^{0.62} & \mbox{for } |\k| > m_g 
\end{array}  \right. , \label{f}
\end{equation}
originates from renormalizing the composite Coulomb kernel. The
gluon mass, $m_g$ arises from dimensional transmutation and can be fixed by
the string tension. The result of the fit to lattice data,
 yields $m_g \sim 1.6/r_0 \sim 600\mbox{ MeV}$ and is show in Fig.~6. 
 The gluon gap function $\omega(|\k|)$ is well approximated by,  
\begin{equation}
\omega(|\k|) = \left\{ \begin{array}{cc} m_g & \mbox{ for } |\k| <
    m_g \\ |\k| & \mbox{ for } |\k| > m_g \end{array}
 \right. . 
\end{equation}
The first two terms in Eq.~(\ref{gap}) are then given by 
\begin{equation}
\delta m_0 =  \delta m_0(|\q|) = |\q| \sin_\q , 
\end{equation}
and
\begin{widetext}
\begin{equation}
 \delta m_C = \delta m_C(|\q|) 
 = - {C_F\over 2}\int {{d\k}\over {(2\pi)^3}} 
V_{eff}(|\k-\q|) \left[ \sin_\k \cos_\q - {\hat \k}\cdot {\hat \q} \sin_\q \cos_\k
\right]. 
\end{equation}
\end{widetext}
\begin{figure}[htb]
\includegraphics[scale=0.5]{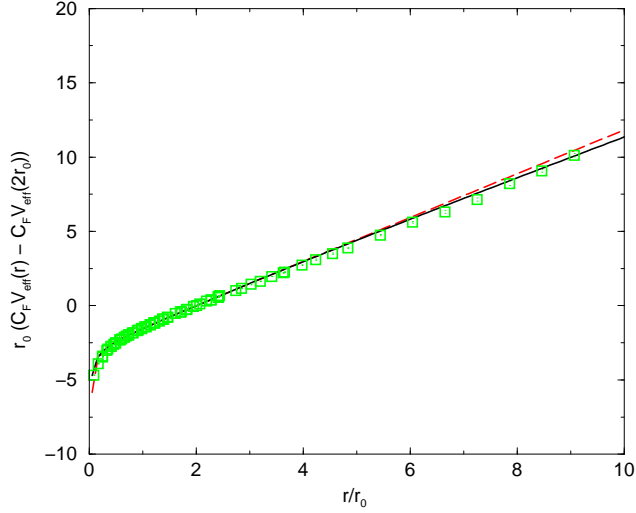} 
\caption{\label{fig:6} Comparison of the lattice results for the ground
  state potential between two static ${\bf 3}$ and $ {\bf \bar 3}$
  sources and the fit to $V_{eff}$ given by Eq.~(\ref{potss}) (solid
  line), and Eq.~(\ref{vc}) (dashed line). Lattice data (boxes) come
  from Ref.~\cite{latt1} }
\end{figure}
The new contribution to the gap arising from 
 the  $Q{\bar Q}G$ cluster contains matrix elements of 
 $H$ evaluated between the BCS vacuum and a tree particle, $Q{\bar
 Q} G$ state or between  $Q{\bar Q} G$ and $Q{\bar Q}$ states. 
 Only $V_{qg}$ and $H_C$ contribute to those and they are 
 of order $O( g\sim
 \langle d(|\k|) \rangle)$ and $O( g^3\sim \langle d^3(|\k|) \rangle)$
 respectively. As discussed in Ref.~\cite{ss7} the later is of a type
 of a vertex correction and is expected  to be a small $O(10 - 20\%)$ 
 correction to an $O(g)$ contribution from $V_{qg}$. Therefore we will 
 not further included it here (this is also 
  consistent with the ring-rainbow approximation to $V_{eff}$).  
 The final expression for $\delta m_g$ also requires $Q{\bar Q} G$
 wave functions {\it i.e} the eigenstates of $H_0 + H_C$ projected
 onto the $Q{\bar Q}G$ states. In this work we do not attempt to solve 
 this eigenvalue  problem instead  we will approximate the sum over 3-particle
 intermediate states by, 
\begin{widetext}
\begin{equation}
\sum_n |n\rangle {1\over {E_n - E_{\Omega_{BCS}}} } \langle n| 
 \longrightarrow   |\Psi \rangle {1\over {E_\Psi - E_{\Omega_{BCS}}}} \langle \Psi | 
 + \sum_{Q{\bar Q} G, (E_{Q{\bar Q}G}-E_{\Omega_{BCS}}) > \Lambda_F} |Q{\bar Q} G \rangle 
{1\over {E_{Q{\bar Q} G} - E_{\Omega_{BCS}}}} \langle Q {\bar Q} G |, 
\end{equation}
\end{widetext}
{\it i.e.} we approximate the sum over the complete set of
eigenstates by a single state with energy smaller then some factorization
 scale, $\Lambda_F$, $E_\Psi - E_{\Omega_{BCS}} < \Lambda_F$,  and a
 perturbative continuum of states with energy 
 grater then $\Lambda_F$.  The scale $\Lambda_F$ should roughly equal the
 energy where, due to string  breaking  the linear confining potential 
 saturates. For the first excited hybrid  potential $\Lambda_F \gsim
 1.5  \mbox{GeV}$ which corresponds to the distance between color sources,
  $r = 1.15\mbox{ fm}$~\cite{Bali}. Thus we expect that the size of the 
 momentum space wave function, $\langle Q{\bar Q} G|\Psi\rangle$ 
  should be of the order $\beta \sim 1/r = 0.2 \mbox{GeV}$. 
 As for the spin-orbital momentum dependence of the $Q{\bar Q}G$ wave
 function we shall assume that it corresponds to low values of the 
 orbital angular momenta which are consistent with those of the low 
 lying gluonic excitations in the presence of $Q{\bar Q}$ sources. 
 Lattice computation of the $Q{\bar 
  Q}$ adiabatic potentials arising from excited gluon configurations
 indicate that the so called $\Pi_u$ potential has lower energy
 then the $\Pi_g$ potential~\cite{latt1}. These two correspond to
 gluon  configuration with $J^{PC}=1^{+-}$ and $1^{--}$ respectively which is
  also consistent with the bag model representation of gluonic 
 excitations~\cite{bag}. The 
 $Q{\bar Q} G$ wave function coupled with the $J^{PC}= 1^{+-}$
  gluon quantum numbers would also have the $Q{\bar Q}$  pair with 
 the same $J^{PC} = 1^{+-}$ quantum numbers  (to give the overall
 $J^{PC} = 0^{++}$ of the vacuum) and would be given by 
\begin{eqnarray}
\left[Q{\bar Q} G\right]^0 & = &  
\left[ \left[ (L_{Q\bar Q}=1) \times (S_{Q\bar Q} = 0)\right]^1
\right. \nonumber \\
& \times  & 
 \left.  \left[ (L_G = 1) \times (S_G = 1) \right]^1 \right]^0. 
\end{eqnarray}
It is easy to check, however, that since $V_{qg}$ is spin dependent
this wave function has vanishing overlap with the 
$V_{qg}|\Omega\rangle$ state. 
 The other possibility is to take 
 $J^{PC} = 1^{--}$ configurations for both the glue and the
 quark-antiquark which give, 
\begin{eqnarray}
 \left[Q{\bar Q} G\right]^0 & = &  
\left[ \left[ (L_{Q\bar Q}=0) \times (S_{Q\bar Q} = 1)\right]^1
\right. \nonumber \\
& \times  &
 \left.   \left[ (L_G = 0) \times (S_G = 1) \right]^1\right]^0 . 
\end{eqnarray}
and take the spin-orbit wave function in the form of, 
\begin{widetext}
\begin{equation}
 \langle Q(\k_Q,\lambda_Q, i_Q),{\bar Q}(\k_{\bar Q},
 \lambda_{\bar Q},i_{\bar Q}), 
 G(\k_G,\lambda_G,a) | \Psi \rangle  
 =  (2\pi)^3\delta(\k_Q + \k_{\bar
 Q} + \k_G) 
U^{\dag}(\k_Q,\lambda_Q) 
  \bbox{\alpha}\cdot \bbox{\epsilon}(\k_G,\lambda_G) V(\k_{\bar
 Q},\lambda_{\bar Q}) \Psi(\k_Q,\k_{\bar
 Q},\k_G). 
\end{equation}
\end{widetext}
 The color part of the wave function is given by 
  $T^a_{i_Q,i_{\bar Q}}$, and for the orbital wave function we
  will take a gaussian ansatz, 
\begin{equation}
\Psi(\k_Q,\k_{\bar Q},\k_G) = exp(-(\k^2_Q + \k^2_{\bar Q} +
 \k^2_G)/\beta^2). \label{orb}
\end{equation}
The expression for $ \delta m_g$ is then given by, 
\begin{equation}
\delta m_g = \delta m_g(|\q|) = \delta m_{g,soft} + \delta m_{g,hard}
, 
\end{equation}
with 
\begin{widetext}
\begin{eqnarray}
\delta m_{g,soft} = - & &  { {C_F}\over {E_\Psi -
 E_\Omega}  }   
 \int {{d\k} \over {(2\pi)^3} }
 {{d(|\k-\q|) \Psi(\k,\q,\k-\q)/|\Psi|} \over {\sqrt{\omega(|\q-\k|)}}}
 \left[ s_\k c_\q - I(\k,\q) c_\k s_\q \right]  \nonumber \\
 & & \times 
 \int {{d\p} \over {(2\pi)^3}} {{d\l}\over {(2\pi)^3}} 
 {{ d(|\l-\p|)\Psi^{*}(\p,\l,\l-\p)/|\Psi|} \over
 {\sqrt{\omega(|\l-\p|)}}}
 \left[ 1 + s_\p s_\l + I(\p,\l) c_\p c_\l \right], 
\end{eqnarray}
\begin{equation}
\delta m_{g,hard} = -  C_F 
 \int {{d\k} \over {(2\pi)^3} }
 { { d^2(|\k-\q|) } \over {|\q-\k|}}
 {{ \left( 1-\Psi_{\Lambda_F}(\k,\q,\k-\q) \right)}
  \over {|\k| + |\q| + |\k-\q|}}
 \left[ s_\k c_\q - I(\k,\q) c_\k s_\q \right]  .
\end{equation}
\end{widetext}
Here
\begin{equation}
I(\k,\q) \equiv  { { (|\k|^2 + |\q|^2) {\hat \k}\cdot {\hat \q}  - 
 |\k||\q|(1 + ({\hat \k}\cdot {\hat \q})^2 ) } 
 \over {(\k - \q)^2} },
\end{equation}
\begin{equation}
|\Psi|^2 =  \int {{d\p} \over {(2\pi)^3}} {{d\l}\over {(2\pi)^3}}
 |\Psi(\p,\l,\l-\p)|^2
 \left[ 1 + s_\p s_\l + I(\p,\l) c_\p c_\l \right]  ,
\end{equation}
and $\Psi_{\Lambda_F}$ given by Eq.~(\ref{orb}) with $\beta \to
 \Lambda_F$ so that $1 - \Psi_{\Lambda_F}$ cuts off hard $Q{\bar Q} G$
 contribution for energies below $\Lambda_F$.

\subsection{ UV behavior and renormalization }

Before analyzing the full gap equation and in particular the 
 effects of $\delta m_g$, we shall first discuss the IR and UV
 behavior in the BCS approximations. 
 The BCS approximation to the chiral gap has been studied earlier 
 for various model approximations to $V_{eff}$. Most of them 
  use an effective  potential which is 
  regular at 
 at the origin,  {\it e.g} a pure linear potential $V_{eff}(r) = b r$~\cite{chha,Cotan} or a
 harmonic oscillator, $V_{eff}(r) = k r^2$~\cite{Lis}. For such potentials the
 gap equation is finite in the high momentum limit and no 
 renormalization is  required. This is not the case if potential
 has the  Coulomb  component with 
 $V_{eff}(r\to 0)  \sim \alpha /r $ and $\alpha$
 being either a constant or a running coupling $\alpha \to
 \alpha(r) \sim 1/\ln(1/r)$. The BCS quark gap  for potentials with
 the  Coulomb tail was studied in Ref.~\cite{AA,swif} and
 Ref.~\cite{ss3}.   The gap equation used in Ref.~\cite{AA} would be 
  identical to one used here, if $\delta m_g$ was set to zero ({\it
  e.g.} the BCS approximation). 
  Instead,  in Ref.~\cite{AA} an energy-independent interaction  
 motivated by a transverse gluon exchange was added. 
  In Ref.~\cite{AA} it was argued  that, in the chiral limit, 
  the resulting gap equation, 
 could be renormalized by introducing a single counterterm
 representing the wave function renormalization. 
  Starting from the Coulomb gauge Hamiltonian this would arise if the free
 quark kinetic energy term was replaced by a renormalized one,  
  \begin{eqnarray}
& & \int d\x {\bar \psi}(\x) \left[ -i \bbox{\alpha}\cdot \bbox{\nabla}
\psi(\x)\right] \nonumber \\
& &  \to Z(\Lambda) 
 \int d\x {\bar \psi}(\x) \left[ -i \bbox{\alpha}\cdot \bbox{\nabla}
 \psi(\x)  \right]_\Lambda
\end{eqnarray}
The explicit, UV  cutoff-$\Lambda$ dependence regularizing the
kinetic operator can be introduced, for example by field smearing,
however, the regularization procedure becomes irrelevant once the resulting
 gap equation is renormalized.  The unrenormalized BCS,  
 gap equation (without effects from transverse
gluons) is then given by, 
 \begin{widetext}
\begin{equation}
Z(\Lambda)m(|\q|) =  {C_F\over 2}\int^\Lambda {{d\k}\over {(2\pi)^3}} 
 V_{eff}(\k-\q) {{ m(\k)}\over {\sqrt{\k^2 + m^2(\k)}}} 
-  {C_F\over 2}\int^\Lambda {{d\k} \over { (2\pi)^3}} V_{eff}(\k-\q) {\hat \k} \cdot {\hat
 \q} { {|\k|} \over {|\q|} } {{m(\q)}\over {\sqrt{\k^2 + m^2(\k)}}},
\label{gap1}
\end{equation}
\end{widetext}
where we have defined the {\it constituent} mass, $m(|\k|)$ by
$m(|\k|)  \equiv |\k|
\sin_\k$. The renormalized equation is obtained by a single
 subtraction {\it i.e.} by fixing the $\Lambda$-independent solution, 
 $m(|\k|)$ at a specific value of $|\q|  = |\q_0|$. This leads to a
 ($\Lambda$ and $\q_0$-independent), renormalized gap equation, 
\begin{widetext}
\begin{equation}
  m(|\q|) \lim_{\Lambda \to \infty} \left[ I_m(|\q_0|,\Lambda) -
    I_Z(|\q_0|,\Lambda) \right] 
 = m(\q_0)\lim_{\Lambda\to \infty}  \left[ I_m(|\q|,\Lambda) 
- I_Z(|\q|,\Lambda) \right] ,
\end{equation}
\begin{equation}
I_m(|\q|,\Lambda) \equiv {C_F\over 2}\int^\Lambda {{d\k}\over {(2\pi)^3}} 
 V_{eff}(\k-\q) {{ m(\k)}\over {\sqrt{\k^2 + m^2(\k)}}} ,\;\;
I_Z(|\q|,\Lambda) \equiv {C_F \over 2}
  \int^\Lambda {{d\k} \over { (2\pi)^3}} V_{eff}(\k-\q) 
 {\hat \k} \cdot {\hat \q}
  { {|\k|} \over {|\q| }} {{m(|\q|)}\over {\sqrt{\k^2 + m^2(\k)}}} , 
\end{equation}
\end{widetext}
Whenever possible we will also use the notation $I(|\q|) 
\equiv I(|\q|,\infty)$.  
 We will now show that, 
  this equation does not have a well behaved, nontrivial 
 solution vanishing asymptotically in the large momentum limit, as it
 was assumed, for example in Ref.~\cite{AA}. Before we do that first we
  need to take care of the possible IR divergences which appear in
  the integrals when $\k \to \q$. In this
  limit $V_{eff}(\k-\q)$ is highly divergent, reflecting the long range
 nature of  the confining interaction, {\it e.g.} $V_{eff}(\k-\q) \propto
 1/(\k-\q)^4$ for the linear potential. The gap equation, however, is 
 finite due to cancellation 
 of the numerators between $I_m$ and
  $I_Z$. To make individual integrals well behaved 
  we can split the IU and UV parts of
 $V_{eff}$ defining, 
\begin{eqnarray}
& & V^{IR}(\k,M) \equiv \theta(M-|\k|) V_{eff}(|\k|), \nonumber \\
& & V^{UV}(\k,M) \equiv \theta(|\k|-M) V_{eff}(|\k|),
\end{eqnarray}
since $V^{IR}(\k,M) + V^{UV}(\k,M) = V_{eff}(\k)$ and gap equation is
independent on the parameter $M$ and we will not write it
explicitly. The gap equation becomes, 
\begin{equation}
 m(|\q|) 
 =  {1\over {A + B(|\q|)}} \left[ I^{IR}_m(|\q|) - I^{IR}_Z(|\q|)  +
 I^{UV}_m(|\q|) \right],    \label{gapuv}
\end{equation}
where
\begin{equation}
 A = A(|\q_0|) = 
 {
{\left[I^{IR}_m(|\q_0|) -
      I^{IR}_Z(|\q_0|) + I^{UV}_m(|\q_0|) \right]} \over {m(|\q_0|)} },
\end{equation}
and 
\begin{equation}
  B(|\q|) =  B(|\q|,|\q_0|)  = {{I_Z^{UV}(|\q|)}\over {m(|\q|)}}
 -  {{I_Z^{UV}(|\q_0|)}\over {m(|\q_o|)}}. \label{BB}
\end{equation}
For given $\q_0$, $A$ is a constant and $B$ is a function of $\q$,
and both, $A$ and $B$  are well defined. In $A$ the IR divergences 
 cancel between $I^{IR}_m(|\q_0|)$ and $I^{IR}_Z(|\q_0|)$, 
 and $I^{UV}_m$ is finite if $m(|\q|) \to 0$ as $\q \to \infty$. 
 In $B$ each term is IR finite and the UV 
   divergences cancel  between the two terms in Eq.~(\ref{BB}).
It is easy to show that for $|\q|>> M,|\q_0|$ the function $B(|\q|)$ , 
 behaves as, 
\begin{equation}
B(|\q|) \to - C_F { \alpha \over {3\pi}} \log \q^2
\end{equation}
 for $V_{eff}(|\k|) \to 4\pi \alpha/|\k|^2$ as $|\k| \to \infty$. If
 $\alpha$ is replaced by a running coupling then $|B(|\q|)|$ 
 grows with $|\q|$ like 
 $\log\log \q^2$. From Eq.~(\ref{gapuv}) it thus follows that 
 for some $|\q|/|\q_0| >> 1$, 
 $A + B(|\q|)$ changes sign and therefore the equation is undefined. 
 This also remains true if an additional
  transverse potential is added as done in ~\cite{AA}. 
   In this case the
 the argument of the integrals defining function the $B(|\q|)$ becomes, 
\begin{eqnarray}
  {\hat \k} \cdot {\hat \q}
  { {|\k|} \over {|\q|}} V^{UV}(\k-\q)  \to 
   { {|\k|} \over {|\q|}} 
 & & \left[ V^{UV}(\k-\q)  {\hat \k} \cdot {\hat \q} \right.  \nonumber \\
 & &  \left.  + 2 I(\k,\q) V^{UV}_T(\k-\q) \right] \nonumber \\
\end{eqnarray}
 For $V_T(\k) = 4\pi \alpha /(\k^2 + const)$ (as used in 
 Ref.~\cite{AA}) the  
 additional transverse potential does not
 contribute to the  $\log\q^2$ (or $\log\log\q^2$) behavior of $B(|\q|)$.
 In our case there would be a similar contribution arising from the
 hard part of  the gluon exchange given by $\delta m_{g,hard}$.  
 At large $|\q|$ it 
 adds, a {\it positive },  $+ C_F {\alpha \over {12\pi}} \log \q^2$
 contribution to $B(|\q|)$ and therefore does not cause problems on
 its own but at the same time does not eliminate the singularity from
 the Coulomb potential since the net effect is such that 
 $B(|\q|) \to - \infty$ as $|\q| \to \infty$. 

 The problems with the renormalized gap equation for
 the  Coulomb potential is illustrated in Figs.~7 and 8. 
 In this test case we simply take 
 \begin{eqnarray}
V_{eff}(\k)  & = &  V^{IR}(\k) + V^{UV}(\k) \nonumber \\
& = & {1\over C_F} {{8 \pi b } \over {\k^4}} + {{4\pi \alpha
    } \over {\k^2 \log( \k^2/m_g^2 + 2)^n}} \label{vc}
\end{eqnarray}
 For the string tension, $b=0.24\mbox{ GeV}^2$, $\alpha = 0.1$, and
$n=0$ this gives a good fit to the lattice data as shown in Fig.~6. 
\begin{figure}[htb]
\includegraphics[scale=0.5]{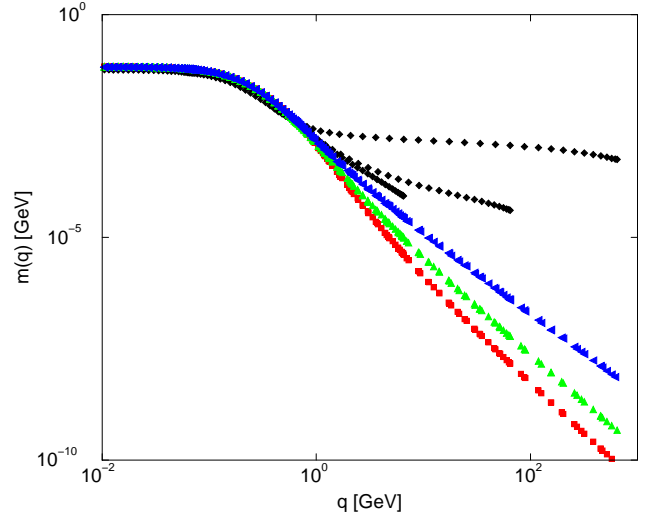} 
\caption{\label{fig:7} Solution of the gap equation for a potential given
  by Eq.~(\ref{vc}), with  $\alpha=1$, and $n=0,1,1/2,3/2$. The three
  upper lines correspond to $n=0$, the next three to $n=1$, $n=1/2$
  and $n=3/2$ respectively } 
\end{figure}
\vskip 20pt 
\begin{figure}[htb]
\includegraphics[scale=0.5]{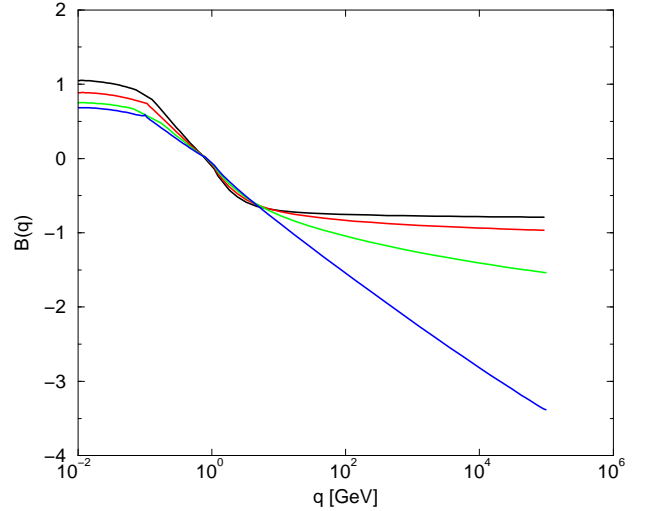} 
\caption{\label{fig:8}  Function $B(|\q|)$ calculated for the potential as in
  Fig.~7. The lines (from top to bottom at high $|\q|$) correspond to $n=3/2$,
  $n=1/2$, $n=1$ and $n=0$ respectively } 
\end{figure}
In Fig.8 we show the function $B(q)$ calculated from a numerical
solution  to the gap equation,
Eq.~(\ref{gapuv}). 
 Since $V_{eff}$ is already given by a
 sum of two terms, one dominating in the IR and the other in the UV 
  the components  $V^{IR}$  and $V^{UV}$ can be defined as the 
  linear and the Coulomb piece respectively. If $n>1$ there is no
  renormalization required and the gap equation is given by Eq.~(\ref{gap1})
 with $Z=1$. The $n=3/2$ case  corresponds to an approximate
 analytical solution for $V_{eff}$ discussed in~\cite{ss7} and is
 also close to the exact, numerical solution given by
 Eqs.~(\ref{potss}).  
  The solution of the gap equation, $m(q)$ for $n=3/2$ is shown in Fig.~7
 by the lowest line (boxes). The function $B(|\q|)$ corresponding to this case is
 shown in Fig.~8 by the upper solid line (at large $|\q|$), 
 which asymptotically approaches $B(|\q|) \to const 
 - 2 C_F \alpha/(3\pi \log(\q^2))$ as $|\q| \to
 \infty$.  We then take this solution to set the value of
 $m(|\q_0|=m_g=600\mbox{ MeV})$ and solve the renormalized gap equation,
 Eq.~(\ref{gapuv}) for $n=1$, $n=1/2$ and $n=0$. The asymptotic behavior
 at large $|\q|$ of $B(|\q|)$ for these three cases is given by, 
 \begin{equation}
B(|\q|) \to \left\{ \begin{array}{cc}  - C_F {\alpha\over {3\pi}}
    \log\log(|\q|^2),  &   n = 1 \\
      - 2 C_F {\alpha \over {3\pi}} \log^{1/2}(|\q|),  & n = 1/2 \\
     - C_F {\alpha \over {3\pi}} \log(|\q|),&  n = 0 
 \end{array} \right. 
\end{equation}
 The corresponding solutions to the gap equation are shown by the 
 five upper lines in Fig.~7. 
 The highest three correspond to
  $n=0$ case and their splitting indicates that the 
 numerical procedure has not converged into a unique solution. 
 These three solutions correspond to three 
  different cut-offs on the maximum momentum, $|\q|_{max}  = 10m_g, 100m_g$ and $1000m_g$. 
 The other two lines correspond to solutions for $n=1/2$ and $n=1$
 respectively. In these two cases 
 the same three values for the 
 momentum cutoffs were used and apparently in both cases a cutoff independent
 solution has emerged. 
  This is because for $n=1/2$ and
  $n=1$  $|B(q)|$  grows very  slowly and in practice the zero 
  of  the denominator in Eq.~(\ref{gapuv}) is not crossed. 
  This test calculation was performed with unphysically large
 $\alpha=1$. For $\alpha \lsim 0.5$ numeral computations,  
 which always have a build in a finite upper momentum cutoff 
 converge for $|\q|_{max}$ as large as $10^6m_g$. 

It is clear that the problematic UV contributions
originate from need for wave function renormalization. This 
problem has been resolved in Ref.~\cite{ss3} using an effective
Hamiltonian with perturbative $O(g^2)$ contributions calculated 
 via a similarity
transformation~\cite{Glazek}. In that approach, in addition to the Coulomb and
transverse gluon contributions to the gap equation, $\delta m_C$ and 
 $\delta m_{g,hard}$,  there was also a 
 modification of the single particle kinetic. The additional
 contribution to  the gap equation via $\delta m_0$ cancels the
  $\log|\q|$ term from $B(|\q|)$ and results in a well defined
  equation.  The disadvantage of that approach
 however, is  that it is restricted to the free rather then BCS
 basis and so far it has not being generalized beyond perturbation
 theory. 

The resummation of the leading UV contribution to the
Faddeev-Popov operator and the Coulomb kernel has the effect of
softening the UV behavior ({\it c.f.} Eqs.~(\ref{d}),~(\ref{f})) and
 at the BCS level leads to a finite gap equation without need 
 for any  additional, {\it e.g.} wave function renormalization
 counter-terms.  Furthermore the  $e^{-S}$ method enables to include 
 effects of  transverse gluons with a well defined energy dependence. 
  The large momentum contribution from the 
  $Q{\bar Q}G$ cluster, $\delta m_{g,hard}$ still requires
 renormalization through the wave function counterterm. As discussed
 above since it leads to $B(|\q|)$ which is positive at large $|\q|$
 the renormalized gap equation is well behaved. 

\subsection{ Numerical results } 

We will now discuss the numerical results.  These are summarized
Fig.~9. 
\begin{figure}[htb]
\includegraphics[scale=0.5]{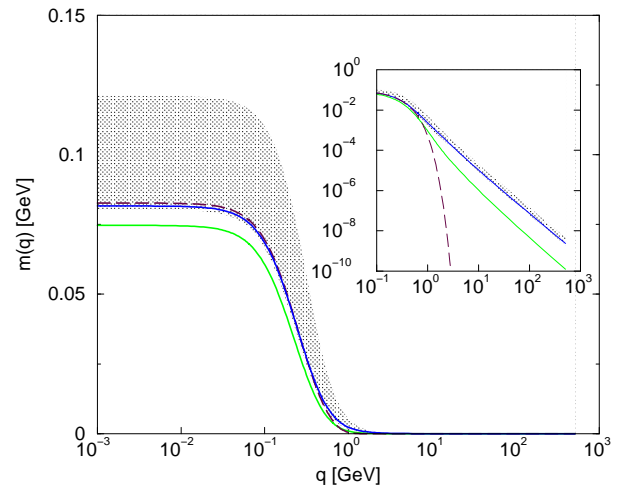}
\caption{\label{fig:9} Solution to gap equation. The dashed line  
  is a solution with the IR part of the potential
 (linear potential) only. The lowest solid line comes from a 
 solution using the full 
 static potential, $\delta m_C$. 
 The next higher solid line includes the static potential and
  the hard gluon,  $\delta m_{g,hard}$, contribution  from the $Q{\bar
  Q} G$ sector calculated for $\Lambda_F = 1.5\mbox{ GeV}$. 
 The shaded region corresponds to the full solution with 
 $0.1 \mbox{ GeV} \le \beta  \le 0.4 \mbox{ GeV}$, and $1 \mbox{ GeV}
  \le \Lambda_F \le 2 \mbox{ GeV}$}
\end{figure}
The BCS, potential contribution to the gap equation, $\delta m_C$ is
split into an IR and UV parts by setting $f(\k)=0=d(\k)=0$ for
$|\k| > m_g$ and $|\k| < m_g$ respectively. For the pure IR potential
(dashed line in Fig.~9) the gap function, $m(q)$ is below $100\mbox { MeV}$ for
low $|\q| < m_g$ and it vanishes rapidly (as $1/|\q|^4$) at high $q$. 
 The addition of the UV  component of the potential,   {\it
  i.e.} the Coulomb tail with the $1/\log(q)^n, n>1$ UV 
 suppression, does not change much the low momentum behavior of
 $m(|\q|)$. It  actually decreases $m(0)$ to about $75\mbox{ MeV}$, 
   by it increases the high momentum
 tail,  overall leading to no change in the $\langle {\bar
  Q} Q \rangle$ condensate which stays at about $-(111\mbox{ MeV})^3$. 

 The $\delta m_{g}$ contribution 
  depends on, $\beta$ which sets the size of the 
 soft wave function, $\Lambda_F$ which divides between the soft and 
  hard one-gluon intermediate states, and $E_\Psi-E_\Omega$ which
 determines the energy of the  low gluonic excitations. 
 As discussed earlier it is reasonable to set $\beta \sim 
   0.2 \mbox{ GeV}$ and $\Lambda_F \sim 1.5 \mbox{ GeV}$. 
  As for the energy of the soft $Q{\bar Q} G$ state we take,  
 $E_\Psi - E_\Omega = m_g$ which we expect to
 be close to the lower bound and would therefore give  
 the upper limit on the  $Q{\bar Q}G$ contribution. The effect of the hard 
 one-gluon-exchange contribution,  defined by $\delta m_{g,hard}$,  is
  to increase $m(q)$ yielding $m(0)\sim 80\mbox{ MeV}$ and 
 enhancing the condensate,  $\langle {\bar Q} Q \rangle = -(150\mbox{
   MeV})^3$.  The solution to the gap equation including $\delta m_C$
  and $\delta m_{g,hard}$ is shown by the second to lowest solid line
  in Fig.~9. 

As mentioned above, the gap equation with 
 $\delta m_{g,hard}$ requires
 renormalization and we have simply set $Z=1$ at $\Lambda=|\q_{max}|$. 
 Using any of the three values of $|\q_{max}|$ given previously 
 no effect on the solution could be observed. This is
 analogous to the test case discussed earlier. 

The full effect of the $Q{\bar Q} G$ sector,
 including the soft contribution, parametrized by $\delta m_{g,soft}$
 with the factorization scale $\beta$ in the range between $0.1 \mbox{
   GeV}$ and $0.4 \mbox{ GeV}$ and $\Lambda_F = 1 - 2 \mbox{ GeV}$ 
  is shown by the shaded region. The lower limit corresponds to
 $\beta=0.1\mbox{ GeV}$ and $\Lambda_F = 2\mbox{ GeV}$ and yields 
 $\langle {\bar Q} Q \rangle = -(140\mbox{
   MeV})^3$; for the upper limit $\beta=0.4\mbox{ GeV}$, 
 $\Lambda_F = 1\mbox{ GeV}$ and  $\langle {\bar Q} Q \rangle = -(190\mbox{ 
   MeV})^3$. The addition of the soft gluon intermediate state 
  brings both, the {\it constituent} quark mass 
 and the condensate significantly closer to 
  phenomenologically acceptable values.  

 An  alternative simple parameterization of the soft gluon contribution
 would be to replace it by an effective local operator, by expanding $\delta
  m_{g, soft}$ in powers of $|\k|/m_g$. The lowest dimension operator
  has the structure 
\begin{equation}
\delta V = - {C\over {m^2_g}} 
\int d{\bf x} \left [ \psi^{\dag}({\bf x}) \alpha^i \psi({\bf x}) 
 \delta_{T,ij}(\bbox{\nabla}_{\x})
 \psi^{\dag}({\bf x}) \alpha^j \psi({\bf x}) \right]_{\beta}
\end{equation}
with $C$ being a dimensional constant, $1/m_g^2$ scale arising from
 the product of  $E_\Psi - E_{\Omega_{BCS}}$ and $\omega(0) = m_g$,
 and the operator being related by the factorization scale
 $\Lambda_F$. The appearance of these to scales is quite natural. Since
 the operator arises through elimination of part of the Fock space the 
 overall scale is given by the excitation energy of the eliminated
 sectors and the  momentum cutoff comes from the spatial extend of the 
  excited state  wave function. 
  Such a simple, local  approximation of the soft $Q{\bar Q} G$
 exchange  was considered previously in 
 Ref.~\cite{prl}  where it was shown that such an operator was 
 indeed relevant to chiral
 symmetry braking effects, {\it e.g.} the condensate and the
 $\pi-\rho$ mass splitting. 

\section{Summary} 
Dynamical breaking of chiral symmetry is the fundamental property of
 QCD which leads to the constituent representation. In the 
 canonical, Hamiltonian based formulation it arises via the BCS-like 
 pairing between light quark and antiquarks mediated by the
 attractive Coulomb interaction. However, the extent of chiral symmetry  
 breaking generated this way, as measured by the scalar quark density
 or as compared to the phenomenological constituent quark model is to
 small. We have shown that the naive inclusion of the short range part
 of the $Q{\bar Q}$ potential leads to instabilities in the quark gap
 equations which cannot be renormalized away. In contrast a 
 systematical re-summation of the leading IR and UV corrections to to
 the bare Coulomb kernel leads to an effective interactions which is
 consistent with the variational treatment and the gap equation. 
 Using the linked cluster expansion we have estimated the
 role of three-particle, $Q{\bar Q}G$ correlations in the vacuum, and
 shown that they are indeed important, in particular their 
 low momentum components. 
 Even though we have not  used the
 exact solution describing the soft $Q{\bar Q}G$ state  our result are 
  expected to be close to the upper bound for the non-BCS contribution
 to the chiral condensate and are consistent with previous 
 studies.

\section{Acknowledgment}
We would like to thank Bogdan Mihaila for discussion of the $exp(S)$ 
 method. This work was supported by the US Department of Energy under contract
DE-FG02-87ER40365.

\end{document}